\documentclass[a4paper,fleqn]{cas-dc}

\usepackage[numbers]{natbib}

\usepackage{graphicx}
\usepackage[colorinlistoftodos]{todonotes}

\usepackage{adjustbox}
\usepackage{array}
\usepackage{booktabs}
\usepackage{caption}
\usepackage{float}
\usepackage{longtable}
\usepackage{multirow}
\usepackage{makecell}
\usepackage{placeins}
\usepackage{stfloats}
\usepackage{subfigure}
\usepackage{tabularx}
\colorlet{color1}{black}
\colorlet{color2}{blue}
\colorlet{color3}{black}

\def\tsc#1{\csdef{#1}{\textsc{\lowercase{#1}}\xspace}}
\tsc{WGM}
\tsc{QE}
\tsc{EP}
\tsc{PMS}
\tsc{BEC}
\tsc{DE}

\begin{document}
\let\WriteBookmarks\relax
\def\floatpagepagefraction{1}
\def\textpagefraction{.001}
\shorttitle{Knowledge Prompting: How Knowledge Engineers Use Generative AI}
\shortauthors{E. Koutsiana et~al.}

\title [mode = title]{Knowledge Prompting: How Knowledge Engineers Use Generative AI}                      



\author[1]{Elisavet Koutsiana}[orcid=0000-0001-6544-0435]
\cormark[1]
\ead{elisavet.koutsiana@kcl.ac.uk}


\affiliation[1]{organization={Department of Informatics, King's College London},
                city={London},
                postcode={WC2B 4BG}, 
                country={United Kingdom}}

\author[1]{Johanna Walker}[orcid=0000-0002-5498-8670]
\cormark[1]
\ead{johanna.walker@kcl.ac.uk}

\author[1]{Michelle Nwachukwu}[orcid=0009-0007-7546-4888]
\ead{michelle.nwachukwu@kcl.ac.uk}



\author[1]{Bohui Zhang}[orcid=0000-0001-5430-1624]
\ead{bohui.zhang@kcl.ac.uk}

\author[1]{Albert Meroño-Peñuela}[orcid=0000-0003-4646-5842]
\ead{albert.merono@kcl.ac.uk}
\ead[URL]{https://www.albertmeronyo.org}

\author[1]{Elena Simperl}[orcid=0000-0003-1722-947X]
\cormark[1]
\ead{elena.simperl@kcl.ac.uk}
\ead[URL]{https://elenasimperl.eu}

\cortext[cor1]{Corresponding author}


\begin{abstract}
Despite many advances in knowledge engineering (KE), challenges remain in areas such as engineering knowledge graphs (KGs) at scale, automating tasks, and keeping pace with evolving domain knowledge.
KE has used NLP demonstrating notable advantages in knowledge-intensive tasks, but the most effective use of generative AI to support knowledge engineers across the KE activities is still in its infancy. 
To explore how generative AI may enhance KE and change existing KE practices, we conducted a multi-method study during a KE hackathon. We investigated participants' views on the use of generative AI, the challenges they face, the skills they may need to integrate generative AI into their practices, and how they use generative AI responsibly.
We found participants felt LLMs could indeed contribute to improving efficiency when engineering KGs, but presented increased challenges around the already complex issues of evaluating KE task success.
We discovered prompting to be a useful but undervalued skill for knowledge engineers working with LLMs, and note that NLP skills may become more relevant across more roles in KE workflows. Integrating generative AI into KE tasks needs to be done with awareness of potential risks and harms. Given the limited ethical training most knowledge engineers receive, solutions such as our proposed `KG Cards' based on Data Cards could be a useful guide for KG construction.
Our findings can support designers of KE AI copilots, KE researchers, and practitioners using advanced AI to develop trustworthy applications, propose new methodologies for KE and operate new technologies responsibly.
\end{abstract}


\begin{highlights}
 \item Large Language Models (LLMs) can enhance efficiency in knowledge engineering tasks like knowledge graph construction, but their effective use requires new skills, such as advanced prompting and responsible AI integration.
 \item The introduction of tools such as KG Cards can address ethical and transparency challenges in LLM-assisted knowledge engineering by guiding practitioners towards safer and more responsible practices.
 \item Generative AI offers promising potential for assisting less experienced users in KE tasks, but more research is needed to mitigate biases and ensure reliability throughout the KG development lifecycle.
\end{highlights}


\maketitle
;
\section{Introduction}

Knowledge engineering (KE) is the process of capturing, representing and maintaining knowledge in a machine-readable way. It involves the development of knowledge-based systems (KBSs), and tools and principles to operationalise these~\cite{studer1998knowledge,debenham2012knowledge}. 
Since Google launched its knowledge graph (KG) over a decade ago, KGs (especially large ones with millions of entities and billions of facts in triples) have become the predominant way to represent, store, and utilise knowledge, offering several advantages across various domains. Recebtly they have been used to enhance pre-trained language models (PLMs)~\cite{weijie2019kbert} and retrieval-augmented generation (RAG) systems~\cite{lewis-et-al-2020-rag} and offer explanations in artificial intelligence (AI) applications~\cite{tiddi-schlobach-2022-kg-xai}. 
The techniques used for engineering these KGs have benefited from development in Natural Language Processing (NLP) techniques across tasks such as knowledge extraction, completion, and inconsistency detection~\cite{schneider-et-al-2022-decade}. More recently, the most advanced branch of NLP, large language models (LLMs), such as GPT models~\cite{brown-et-al-2020-gpt-3,openai-2023-gpt-4} and the LLaMA series~\cite{touvron2023llama,touvron2023llama2}, have demonstrated notable potential for knowledge-intensive tasks, with applications for KG construction through methods such as knowledge probing \cite{petroni-et-al-2019-lama, alivanistos-et-al-2023, zhang-et-al-2023-llmke}. The integration of KGs and LLMs has become a growing trend that fosters a mutually beneficial relationship: LLMs enhance KG construction and maintenance, while KGs are used to train, prompt, augment, and evaluate LLMs~\cite{pan-et-al-2023-llm-kg}. 

Although research efforts to unify LLM and KG have gained popularity, debates and concerns continue. 
Critical perspectives around misinformation, ethical issues, privacy and security, bias, transparency, accountability and many others have emerged at a broader level and apply to LLM-based systems~\cite{pan2023risk, peng2024securing}. Regardless of improvements in automated processes, it may be that KE will always require some form of human-in-the-loop participation to build trust with end-users and stakeholders~\cite{groth2023knowledge}.

Despite the increasing adoption of LLMs in KE, there is a notable gap in user studies focusing on the user-centric aspects of LLM technology in KE.To explore the opportunities and transformations in KE practices enabled by generative AI, it is essential to analyse how researchers and knowledge engineers interact with this emerging technology in standard KE tasks.  To this end, the idea of a hackathon focused on LLMs in KE originated during the Dagstuhl Seminar on ``Knowledge Graphs and their role in the Knowledge Engineering of the 21st Century'' held in September 2022~\cite{groth2023knowledge}.\footnote{\url{https://www.dagstuhl.de/22372}}. Building on this idea, we organised a four-day hackathon, inviting researchers and practitioners from the AI and KE communities to explore generative AI's potential in automating KE tasks.\footnote{\url{https://king-s-knowledge-graph-lab.github.io/knowledge-prompting-hackathon/}}

The user study presented in this work leverages the hackathon as a test bed. Our investigation included an ethnographic study conducted during the event, complemented by post-event semi-structured interviews and a review of the hackathon's documentary outputs. During the four-day workshop, we observed knowledge engineers working on KE research tasks using generative AI models. 
\textcolor{color3}{KE is a broad field involving numerous tasks, tools and practices. In this study,
we aim address a set of curated topics such as knowledge graphs and their construction with humanin-the-loop.}
Specifically, we examined the obstacles that participants encountered, the skills they needed, and the safety testing practices they used. Our research questions are: 
\begin{enumerate}
\item What are the main challenges experienced by knowledge engineers when using generative AI for KE tasks?
\item How do knowledge engineers evaluate generative AI output for their practices?
\item What skills do knowledge engineers need to incorporate generative AI into their practice?
\item How aware are knowledge engineers of using generative AI responsibly?
\item What factors may affect knowledge engineers' trust and uptake of generative AI technology?
\end{enumerate}


We contribute one of the first user studies on the interaction of knowledge engineers with generative AI. Additionally, we identify key areas of strength and challenge for KE in the era of generative AI. Finally, we propose methods to improve transparency and explainability, including Data Cards for KGs and Model Cards for KG embedding models. \textcolor{color1}{It is worth noting that this study was conducted in 2023, and to this day, there is a rapid evolution in the capabilities of generative AI. However, most of the issues raised by participants in the hackathon have not been satisfactorily resolved in the intervening period, and persist as issues that need addressing.} 
Our findings give direction to designers of KE AI copilots to create responsible applications, guide KE researchers and practitioners in developing new methodologies, and advise practitioners wishing to use advanced AI technologies responsibly.

\section{Background and Related Works}
\label{sec:background}

\subsection{Knowledge Engineering}
\label{sec:ke}
 
Knowledge engineering, the branch of AI concerned with building and managing knowledge-based systems~\cite{schreiber-2000-ke, studer-et-al-1998-ke}, has changed dramatically with the latest innovations in machine learning (ML), NLP, and computer vision. The process of constructing a KG usually involves acquiring knowledge, processing it, and deploying the knowledge graph~\cite{fensel-et-al-2020-kg, tamavsauskaite2023defining}. And yet, as the most recent advances in NLP (especially LLMs) and generative AI demonstrate, questions of how to capture and encode domain knowledge in computational representation remain as challenging as ever~\cite{sarker-et-al-2021-nesy}.

\begin{figure*}
  \centering
  \includegraphics[width=0.99\linewidth]{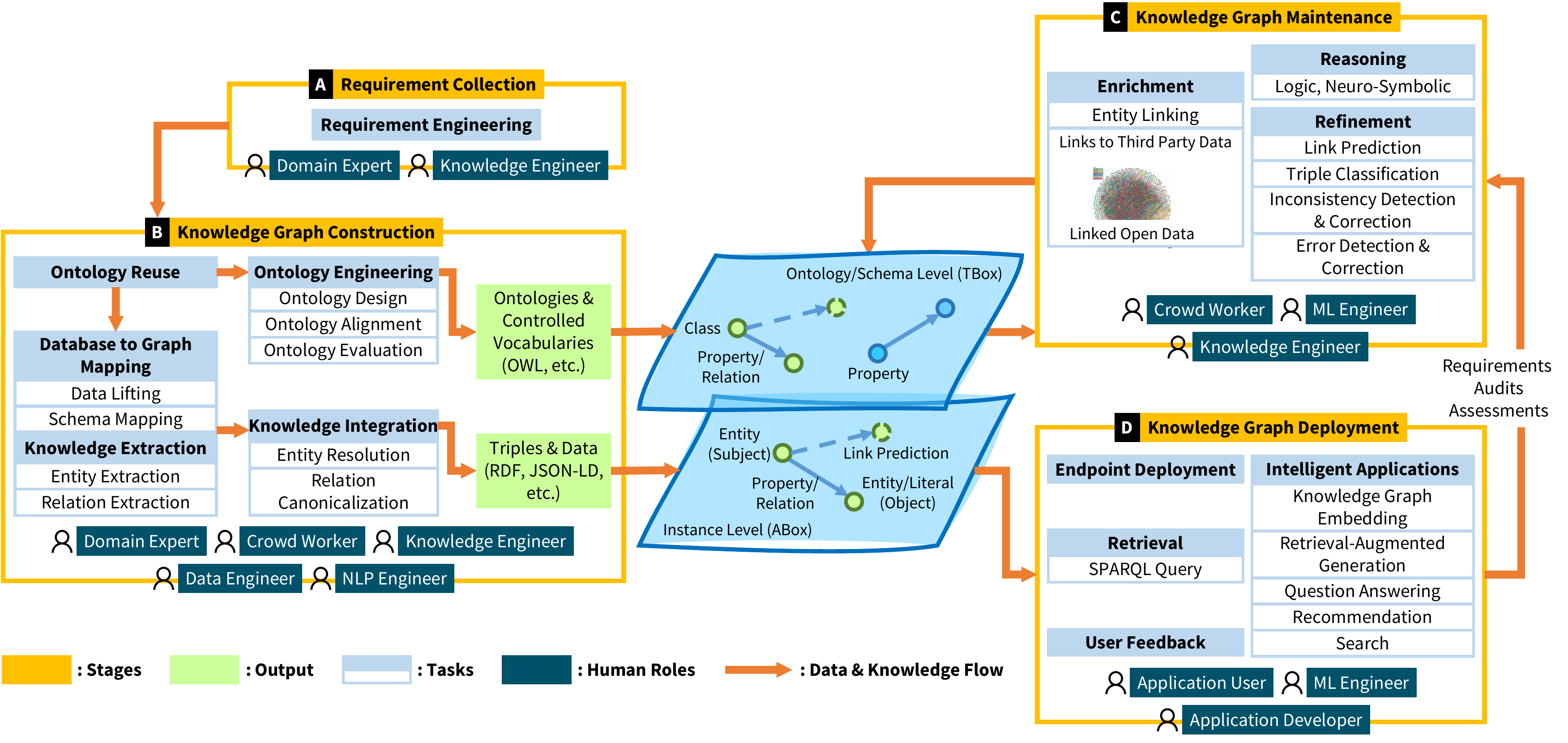}
  \caption{The KG lifecycle today, illustrating the stages from requirement collection to deployment. Each phase --- requirement elicitation, KG construction, maintenance, and deployment --- involves collaboration across roles like domain experts, knowledge engineers, and ML engineers to build, enrich, and deploy KGs for intelligent applications.}
  \label{fig:kg-lifecycle}
\end{figure*}

Figure~\ref{fig:kg-lifecycle} shows that the current KG lifecycle consists of four stages.  This constitutes a mix of automated and manual capabilities and contributions from several stakeholder groups: knowledge engineers, ML specialists, subject domain experts, online volunteers, and crowdsourcing services, as well as developers of applications using KGs. As the figure suggests, \textbf{KGs are interacting with AI capabilities in complex ways}. Human-in-the-loop tasks in the KG lifecycle increasingly use ML models with varying levels of interpretability.

On the left side of the figure, at stage \textbf{A}, which is an entry point and essential step of the KG lifecycle, knowledge engineers (e.g., ontologists) and KG stakeholders (e.g., domain experts) first determine the scope of work and the success criteria~\cite{kendall-mcguinness-2019-requirements}. With the support of description logics, knowledge formalised within a well-structured and expressive TBox that contains terminological axioms can define the relevant vocabulary for a specific application domain. This process, known as ontology engineering (OE), is primarily manual or semi-automatic~\cite{baader2017introduction}. Ontology requirement engineering, which serves as the starting point for OE, typically involves writing user stories and deriving competency questions (CQs) from these stories~\cite{kendall2019ontology}.

KG construction generally involves extracting knowledge from multiple heterogeneous data sources, ranging from structured and semi-structured to unstructured data, and integrating it into KGs~\cite{tamavsauskaite2023defining}. At stage \textbf{B}, knowledge engineers and other specialists may reuse standard or novel ontologies to build KGs through data lifting and knowledge extraction. Data from multiple sources is transformed into KGs using ML techniques for tasks such as named entity recognition \cite{yadav-bethard-2019-ner}, relation extraction \cite{lin-et-al-2016-nre} and entity reconciliation \cite{sevgili-et-al-2020-nel}. Core KE tasks at this stage include knowledge acquisition and representation. Technologies and end-user tools supporting knowledge acquisition have advanced significantly to meet the scale requirements of modern KGs and lead a growing trend toward adoption of end-to-end approaches \cite{schneider-et-al-2022-decade, ye-et-al-2022-generative-kgc}, which we discuss further in section~\ref{sec:automation}. However, the most effective methods for knowledge representation still rely on human oversight at various levels \cite{simperl-luczak-2014-ontology, simsek-et-al-2022}. Increasingly, human input is focused on augmenting or validating algorithmic suggestions \cite{tamavsauskaite2023defining}. 

KGs can also be created on a larger scale through human collaboration, utilizing crowdsourcing and collaborative-editing platforms \cite{hogan2021knowledge}. Crowd workers and editors have important roles throughout the KG lifecycle, especially in knowledge representation and updates, where annotation tasks such as quizzes and voting are often designed to harness their background knowledge~\cite{acosta-et-al-2013-crowdsource,revenko-et-al-2018-crowdsource,kou-et-al-2022-crowdgraph}. 
Although KGs developed using these methods may face quality issues, including errors~\cite{piscopo-simperl-2019-quality,shenoy-et-al-2021-quality}, disagreements~\cite{koutsiana-et-al-2023-disagree}, and bias~\cite{hogan2021knowledge}, similar transparency challenges exist in crowdsourcing for supervised ML. These challenges stem from the transparency limitations of the digital services commonly used for such tasks, such as Prolific and Amazon's Mechanical Turk, which are black-box, proprietary platforms, making results difficult to replicate or reproduce \cite{qarout-et-al-2019}. Furthermore, educating crowd workers to perform annotation tasks effectively is a non-trivial task~\cite{revenko-et-al-2018-crowdsource}. 

On the right-hand side of the figure, KG maintenance (stage \textbf{C}) is prompted by source updates from stage B, and requirements, audits, and assessments from stage D. To further increase their completeness, correctness, and utility, KGs are refined by completion tasks such as link prediction~\cite{rossi-et-al-2020-embedding-lp} and error detection and correction tasks, etc~\cite{paulheim-cimiano-2017-refinement}.
Quality assessment of KGs is essential following construction and enrichment stages to ensure they are fit for specific purposes. Key dimensions for evaluation include accuracy, coverage, consistency, completeness and understandability \cite{hogan2021knowledge}. Accuracy ensures that data represents real-world entities~\cite{zaveri2016quality}, while completeness and coverage assess how well the KG represents all domain-relevant information~\cite{darari2018completeness}. Validity assessment is supported by constraints expressed in shape languages, including SHACL and ShEx~\cite{gayo2017validating}. By measuring whether the graph conforms to the constraints, knowledge engineers can prevent issues where essential parts of the graph are missing or contradictory. Additional dimensions like coherency, conciseness, and understandability support effective usage by minimizing redundancy and enhancing interpretability, all contributing to the graph’s overall utility for applications~\cite{zaveri2016quality}.

At stage \textbf{D}, KGs support a variety of use cases. They are utilised for querying and reasoning in applications such as search~\cite{wiegmann-et-al-2022-lm-search}, question answering~\cite{chowdhery-et-al-2022-palm,guo-et-al-2022-img2prompt-vqa}, and retrieval-augmented generation~\cite{lewis-et-al-2020-rag,gao-et-al-2023-rag}. Information can be derived from KGs through deductive (e.g., logical rules) or inductive methods (e.g., continuous graph embeddings)~\cite{hogan2021knowledge}. To ensure trustworthiness and legal compliance, both approaches must be transparent and accountable to users~\cite{kelpie,bianchi-et-al-2020-kge-xai}. 

The roles and required skills in ontology and knowledge engineering have evolved significantly. Beyond technical expertise, knowledge engineers are expected to possess strong communication skills, conceptual thinking, and organizational abilities~\cite{mcgraw1989knowledge,mykytyn1994knowledge}. In a collaborative ontology engineering, each team member can play multiple roles, depending on the types of contributions and the technology used \cite{simperl2014collaborative}.
To actively involve domain experts in ontology development, technologies such as controlled natural language, semantic wikis, intelligent user interfaces and social computing are proposed~\cite{denaux2011supporting}. However, the rise of generative AI is reshaping these practices. Collaboration between humans and AI agents is redefining ontology and knowledge engineering, streamlining workflows, and altering the responsibilities of knowledge engineers~\cite{allen2023knowledge}. 

\subsection{Automation in Knowledge Engineering}
\label{sec:automation}


Automated methods have long played critical roles in KE, with NLP supporting tasks ranging from knowledge acquisition to completion. The emergence of LLMs has further propelled automated methods into the mainstream, leading to substantial advancements~\cite{yin2022survey}. However, this progress also introduces new challenges in human-AI collaboration and quality assurance. 

Once the data source is determined, the construction of KGs typically begins with information extraction, which involves identifying key terms --- entities and relations. The task of identifying entities is referred to as \textit{named entity recognition} (NER). A named entity is a word or phrase that is consistently standard for a real-world thing, e.g., an organization, person, location, etc. The task can be either formulated as a sequence labelling or a span classification task. 
Current automated NER methods are deep learning-based and consist of input representation at different levels (character, word, hybrid), context encoders that adopt LSTM and Transformers, and tag decoders using models such as conditional random field (CRF) \cite{li-et-al-2018-dl-ner}. 

After determining entities, the next step is to extract relations that link two entities, which is called \textit{relation extraction} (RE). The task can also be formulated as a span classification task. Given the corpus and a pair of entity strings, the task aims to extract a relation between the entities. Depending on the context given, RE models can be performed at sentence-level, dialogue-level, and document-level. Distant supervision approaches are dominant in current RE tasks, which combine the advantages of prior semi-supervised and unsupervised approaches and leverage a vast amount of semi-structured data stored in a KG~\cite{mintz-et-al-2009-distant, smirnova-mauroux-2018-re-ds}.

The knowledge extracted in previous steps may contain duplicates, requiring further cleaning and integration. The next step, \textit{entity linking} (also known as entity alignment, resolution, or matching), is a crucial step of data and knowledge integration. Given two data sources $A$ and $B$, each containing entities with a shared set of attributes, entity linking aims to identify pairs $(e_1, e_2)$, where $e_1 \in A$ and $e_2 \in B$, such that both refer to the same real-world object. Due to the potential quadratic complexity of matching ($|A| \times |B|$), a pre-processing step known as blocking \cite{papadakis-et-al-2021-blocking} is often used to reduce the complexity. According to Shen et al. \cite{shen-et-al-2015-el-kb}, an entity linking system typically consists of three modules: candidate entity generation, candidate entity ranking, and unlinkable mention prediction. Furthermore, PLMs have demonstrated state-of-the-art performance on large-scale, real-world entity linking tasks~\cite{li2020deep, steiner2024fine}.

The third category is knowledge completion. Most KGs are inherently incomplete, necessitating techniques such as \textit{link prediction} (also referred to as KG completion or reasoning) to infer missing information. Specifically, given a query $q$ with either missing tail entities $(h, r, ?)$ or missing relations $(h, ?, t)$, the task is to complete the query by predicting the missing element. Link prediction models are broadly categorised into two main types: rule-based models, for instance, AnyBURL \cite{anyburl}, and embedding-based models. Embedding-based models, which represent graph elements in continuous vector spaces, are further divided into conventional approaches (e.g, TransE \cite{bordes-et-al-2013-transe}, RESCAL \cite{nickel-et-al-2011-rescal}, NTN~\cite{socher-et-al-2013-ntn}, ConvE \cite{dettmers-et-al-2018-conve}) and graph neural network (GNN)-based approaches (e.g., R-GCN \cite{schlichtkrull-et-al-2018-r-gcn}, KBGAT \cite{nathani-et-al-2019-kbgat})~\cite{rossi-et-al-2020-embedding-lp, zamini-et-al-2022-kg-completion}.

The ability of LLMs to perform knowledge-intensive tasks has been extensively studied, including KG construction and completion~\cite{zhong2021factual,liang2022holistic,peng-et-al-2022-copen}. 
One of the foundational studies, LAMA, examined KG completion by probing language models to extract facts through cloze-style (missing word completion) prompts~\cite{petroni-et-al-2019-lama}. Follow-up work after LAMA has shown improvements in performance~\cite{qin-eisner-2021, zhong2021factual}. TKGCon leverages LLMs to generate high-quality KGs from theme-specific corpora~\cite{ding2024automated}.
Veseli et al. \cite{veseli-et-al-2023} conducted a systematic analysis of the capabilities of language models for automated KG completion. Their findings suggest that LLMs can predict facts with high precision for certain relations in Wikidata, though this is not yet universally applicable. Concurrently, prompt engineering techniques have emerged as a focal point in efforts to elicit knowledge from language models effectively \cite{alivanistos-et-al-2023}, further highlighting the evolving role of LLMs in KG-related tasks.

\subsection{Responsible AI}
\label{sec:ethics}

Since AI systems are \textcolor{color1}{frequently} applied in high-stakes contexts such as the medical domain, finance, government, and education, trust concerns have been raised about the potential harm and risks of opaque AI systems to users and stakeholders, especially in deep learning models that are notoriously difficult to interpret~\cite{adadi-berrada-2018-xai}. 
Integrating generative AI into the KG lifecycle may exacerbate existing issues of transparency, explainability, and bias throughout the development and use chain~\cite{kraft2022lifecycle}. This underscores the need for a thorough understanding of how to use such technology responsibly. Within the scope of this study, the principles of responsible AI most relevant to KE practitioners include transparency, explainability, fairness, and privacy.

\textit{Transparency} refers to the degree to which information about an AI system and its outputs is accessible to individuals interacting with the system --- regardless of whether they are aware of doing so. It seeks to answer the question of ``what happened'' in the system. Effective transparency ensures that appropriate levels of information are available at different stages of the AI lifecycle, tailored to the roles and expertise of the stakeholders engaging with the system~\cite{ai2023artificial}.

Despite its importance, ensuring transparency in the use of generative AI presents significant challenges. 
Studies indicate that LLMs lack sufficient accuracy and reliability to operate without human oversight, particularly regarding the lack of provenance and propensity to hallucinate~\cite{ji2023survey}. 
The scale, probabilistic, and inherent black-box nature of LLMs obscure their internal logic, making their behaviour unpredictable and limitations difficult for users to anticipate~\cite{guidotti-et-al-2018-survey, zhao-et-al-2024-xai-llm}. Additionally, the opacity regarding which data some of the LLMs are trained on not only reduces explainability but means accountability is difficult to establish and increases the risk of the inclusion of personal data and even the reproduction of such data in outputs \cite{nasr2023scalable}. Given that the training data is obtained from large-scale web crawls~\cite{touvron2023llama,touvron2023llama2}, LLMs often reproduce and amplify the unfairness and biases inherent in these sources~\cite{morales2024dsl,zhu2024quite}.

To address these challenges, several methods have been proposed that emphasise comprehensive reporting of model information, including model details, metrics, evaluation and training data, etc. 
Mitchell et al. proposed Model Cards, a framework for transparent model reporting that includes metadata on model details, intended use, factors, demographics, performance, evaluation data, training data, supplementary analysis and ethical considerations \cite{mitchell2019model}. It can be tailored depending on the context and the needs of stakeholders, with additional sections such as interpretability methods~.
Similarly, Pushkarna et al. introduced Data Cards for the transparent and human-centred documentation of ML datasets~\cite{pushkarna2022data}. A Data Card is a structured collection of summaries of dataset metadata, such as attributes, intended use cases, provenance, along with explanations and rationales. With their high degree of interpretive flexibility, Data Cards can act as boundary objects~\cite{star-1989-boundary}, enabling stakeholders to collaborate by analysing data using shared dimensions and a common vocabulary. Specifically for NLP systems, \cite{bender2018data} suggested Data Statements to provide context about datasets, specifically to make NLP users of the dataset aware of issues related to exclusion and biases in the data, or that might emerge with the use of the data. 

Achieving transparency in ML models can also be achieved through \textit{explainability}. eXplainable AI (XAI) encompasses multiple notions, including interpretability, responsibility, and accountability~\cite{adadi-berrada-2018-xai}. According to DARPA's XAI program\footnote{\url{https://www.darpa.mil/program/explainable-artificial-intelligence}}, XAI systems are designed to explain their rationale, characterize their strengths and weaknesses, and help human users to understand ``how'' they will behave in the future. 

Several algorithms and models have been predominently used for XAI. Bach et al. proposed layer-wise relevance propagation (LRP), which decomposes non-linear classifiers like neural networks into several layers of computation and computes relevance scores at each layer that can be used for image classification~\cite{bach-et-al-2015-lrp}. LIME, proposed by Ribeiro et al., learns an interpretable model locally around the prediction~\cite{ribeiro-et-al-2016-lime}. Lundbreg and Lee proposed a unified framework that uses SHAP values to measure the feature importance~\cite{lundberg-and-lee-2017-shap}.
Certain explainers target specific types of deep learning models. Ying et al. developed GNNExplainer \cite{ying-et-al-2019-gnn-explainer}, which identifies subgraphs and relevant node features through counterfactual-based importance measurement, applicable across GNN architectures and graph-based ML tasks. 
With the rise of attention mechanisms \cite{vaswani-et-al-2017-attention}, attention values are increasingly used for explanations. However, debates persist on their adequacy, with critiques arguing that attention is not inherently explanatory \cite{jain-et-al-2019-attention-not} and that more effective alternatives exist \cite{bastings-filippova-2020-elephant}.

In the context of this study, Danilevsky et al. surveyed state-of-the-art XAI models in NLP~\cite{danilevsky-et-al-2020-xai-for-nlp}. Tiddi and Schlobach's systematic literature review \cite{tiddi-schlobach-2022-kg-xai} focused on the integration of KGs into explainable machine learning, where KGs are used as domain knowledge for explanations. Zhang et al. conducted the first examination of XAI methods for KG construction~\cite{zhang2023towards}.
Beyond the technical scope, Miller et al explore XAI from a sociotechnical perspective, drawing insights from philosophy, cognitive science, and social psychology~\cite{miller-2019-xai-social-science}. 

\textit{Fairness} in AI is a key focus in this study, addressing concerns of equality and equity by tackling issues such as harmful bias and unfair outcomes that discriminate against certain demographic groups~\cite{ai2023artificial, ntoutsi2020bias}.
Friedman and Nissenbaum defined bias to refer to computer systems that systematically and unfairly discriminate against certain individuals or groups of individuals in favour of others~\cite{friedman-nissenbaum-1996-bias}. 
Bias exists in various forms, including gender, representatives, and selectivity, and may be encoded into KGs through automated processes. Given that LLMs can generate stereotypical bias, their use in automated KE tasks deserves special attention~\cite{nadeem2021stereoset}.

For example, bias arises in information extraction (IE), NER, and semantic role labelling (SRL). In IE, spurious correlations between entities and classes can skew results, as models often rely on biased statistical dependencies due to unbalanced data distributions~\cite{nan2021uncovering, ghaddar2021context}. 
NER is biased in terms of name regularity and demographics, influenced by biased dictionary definitions or neural network-based extraction methods that exclude protective variables~\cite{ghaddar2021context,zhang2021biasing,mishra2020assessing,paparidis2021towards}. 
SRL, which analyses predicate-argument structures for input sentences, also exhibits demographic biases, such as gender and race bias~\cite{zhao2017men}, and commonsense biases, failing to account for basic commonsense knowledge~\cite{lent2021common}. 
However, mitigating bias is challenging, as humans are naturally prone to various biases~\cite{keidar2021towards,kraft2022lifecycle}. 
Technical approaches to addressing bias include causal inference, which modifies target entities \cite{zeng2020counterfactual, wang2020identifying, wang2021robustness}, and counterfactual IE (CFIE) \cite{nan2021uncovering}, which incorporates structural information to capture non-local interactions critical for IE tasks~\cite{zhang2018graph,jie2019dependency}.

Beyond the above mentioned principles, Responsible AI encompasses several other key themes~\cite{mikalef2022thinking}. 
\textit{Safety} in AI focuses on preventing harm, which includes various themes such as robustness against adversarial attacks, malicious use, reliability and reproducibility, etc~\cite{kazim2021high}. AI safety can describe harms in multiple ways, such as the existential safety risk of the fast development of AI without regulation \cite{bucknall2022current}, and the safety risks of current AI systems, which may undermine trust in their deployment~\cite{shneiderman2020human}. This study specifically focuses on the latter.
\textit{Privacy} ensures data protection throughout the lifecycle of AI systems, both in terms of securing personal information and safeguarding individuals from unwanted exposure~\cite{elliott2022ai}.

\section{Methodology}

During August 2023 we organised a four-day research hackathon for knowledge engineers and AI researchers to investigate KE with prompt engineering.\footnote{\url{https://king-s-knowledge-graph-lab.github.io/knowledge-prompting-hackathon/}} 
The hackathon format offered a tool to explore emerging practices as KE processes and approaches are disrupted by generative AI. The project topics (see Table \ref{tab:Hack-topics}) were defined in a community process with 30 experts \cite{groth2023knowledge}.
The hackathon hosted 39 participants from 15 different institutes and various environments from academia and industry, with 31 PhD students, 2 postdocs, 1 lecturers, 2 professors, and 3 industry members. The hackathon attendees were selected based on their background and experience with KE. They were from European research labs that have a strong profile in either publishing in KE venues, or in offering well-used KE industry products.
During the hackathon, participants are divided into 7 groups of 5 to 7 members, based on their prior experience in KE. Each group explored one of the topics listed in Table~\ref{tab:Hack-topics}. While participants were provided with high-level project descriptions, they had the flexibility to select the tools, methodologies, domains, and LLM techniques they wanted to focus on.

\begin{table*}[h]
  \caption{Overview of hackathon topics with associated tools and methodologies. Detailed descriptions and references for the tools and methodologies listed here can be found in Appendix~\ref{sec:appenC} and Table~\ref{tab:tools}.}
  \label{tab:Hack-topics}
\begin{tabularx}{\textwidth}{>{\hsize=0.8\hsize}X>{\hsize=0.75\hsize}X>{\hsize=.3\hsize}X>{\hsize=.3\hsize}X}
\toprule
\makecell[c]{\textbf{Hackathon Topics}} & \makecell[c]{\textbf{\textcolor{color3}{KE task}} }    & \textbf{KE Tools \& Methodologies}      & \textbf{GenAI Tools \& Methodologies}            
\\ \midrule
Determine if generative AI can extract knowledge structures, including inference rules, to go with facts for KG construction
& \textcolor{color3}{Use prompting to fill in triples, for example Coldplay - BandHasMember - ?}
& Triples investigation                                                         & ChatGPT, Few-shot prompt                          \\ \hline
Create a framework providing tools for collaborative human-AI ontology engineering and requirements elicitation
& \textcolor{color3}{If there is a user story, use LLMs to create competency questions and then build an ontology}
& Competency Questions, eXtreme Design methodology                                            & ChatGPT, Multiple prompting techniques  \\ \hline
Determine how KE tasks can be supported by generative AI 
& \textcolor{color3}{Use ChatGPT to re-create the wine ontology}
& NeOn methodology, Competency Questions, HermiT Reasoner, OOPS! & PaLM, Llama, ChatGPT, Few-shot prompt                          \\ \hline
Determine if generative AI perform reasoning tasks completely in natural language 
& \textcolor{color3}{Develop an agent (i.e. LLM) to collaboratively create a comprehensive knowledge graph with a domain expert. This involves crafting both the graph's structure and content interactively, with the agent taking the lead}
& Triples investigation                                                        & ChatGPT, Multiple prompting techniques                           \\ \hline
Determine if generative AI can perform ontology alignment (i.e. identify and match entities between ontologies)  
&\textcolor{color3}{Given a request to an LLM, explore whether two items in WIkidata should be aligned}

& OAEI                                                  & ChatGPT, Zero-shot prompt                           \\ \hline
Determine if generative AI can be used towards the construction of multimodal KGs   
&\textcolor{color3}{Given a painting, explore whether LLMs can fill in a KG with information, such as painting information, painter's image, painter's information, art historian}
& Investigate triples                                                & mPLUG-Owl, InstructBLIP, Text only prompt (no text + image)         \\ \hline
Determine if we can perform ontology refinement (i.e. techniques for KG completion and correction) using generative AI
&\textcolor{color3}{Use OntoClean within an iterative process to improve an ontology and to better represent knowledge}
& OntoClean                                                       & ChatGPT, Llama, Claud, Few-shot prompt                           \\ \hline
\end{tabularx}
\end{table*}

\subsection{Data Collection and Preparation}

Data was collected using three approaches to ensure comprehensive coverage. A detailed list of datasets is provided in Table~\ref{tab:dataset}. 

\begin{table*}[h]
\caption{Number of documents and descriptions for each category of documents in the dataset.}
\label{tab:dataset}
\begin{tabular}{ll}
\toprule
\makecell[c]{\textbf{Selected Method}} &
  \makecell[c]{\textbf{Collected Documents}} \\ \midrule
Ethnographic study &
  2 sets of observation notes from two of the authors \\
Hackathon reports &
  7 reports from the Hackathon groups \\
Hackathon presentations &
  7 presentations and slides from the Hackathon groups \\
Interviews &
 14 semi-structured interviews with Hackathon participants \\ \hline
\end{tabular}
\end{table*}

\paragraph{1. Ethnographic observation} Two researchers performed an ethnographic observation study, using the `observer as participant' technique for ethnographic research \cite{ethnography}. 
The observer as participant role moves the observer into the study environment and closer to the activity of interest. The participants are aware of the observer, but the observer is not engaging with the participants and does not ask or interact with anyone. 
The researchers observed the various groups and kept detailed notes of the 7 groups' interactions and decisions. 

\paragraph{2. Documents collected from the hackathon} Research output in the form of reports and documentation from the hackathon was collected. Hackathon groups produced reports regarding their experience answering specific questions and slide presentations with their strategies and results. This resulted in 7 presentations and reports.

These two data collection activities were conducted contemporaneously with the Hackathon.

\paragraph{3. Semi-structured interviews} After the hackathon, we contacted 14 participants who had volunteered to participate in follow-up interviews to talk about their opinions and experiences interacting with generative AI for KE tasks. Interviewees had backgrounds in KG construction, KG explainability, ontologies, and machine learning, as well as biology and bioinformatics, data management, FAIR data \cite{Dodds_2020}, and digital humanities. 
We conducted these interviews virtually between 17 August 2023 and 6 September 2023 using Microsoft Teams.\footnote{\url{https://www.microsoft.com/en-gb/microsoft-teams/group-chat-software}} 
The interview guide can be found in Appendix \ref{sec:appenA}. 

We used Otter.ai\footnote{\url{https://otter.ai/}} to transcribe the interview recordings. The transcribed data sources were then uploaded into NVIVO\footnote{\url{https://help-nv.qsrinternational.com/20/win/Content/about-nvivo/about-nvivo.htm}}.

\subsection{Data Analysis}

To address our research questions, we conducted thematic analysis across all the data. We first used an inductive method to search for key themes from the interview topic guide (see Appendix \ref{sec:appenA}). Table \ref{tab:main-themes} shows the main themes identified.
We then used a deductive method. We read the data corpus for grounded themes that emerged in the data. The final codebook is shown in Appendix \ref{sec:appenB}.

This analysis was performed by two researchers, who first read the interview topic guide and together identified the key themes. Then they read the data corpus for the ground themes. Following this, the two researchers, advised by another senior researcher, discussed and confirmed the codebook presented in Appendix \ref{sec:appenB}.

\begin{table*}
\caption{Themes identified through the inductive analysis.}
\label{tab:main-themes}
\begin{tabular}{ll}
\toprule
\makecell[c]{\textbf{Themes}}       & \makecell[c]{\textbf{Description}}                                            \\ \midrule
Background experience & The participants' experience prior to the hackathon             \\ 
Evaluation            & Evaluation issues and suggestions for generative AI outputs                \\ 
Skills                   & The skills and qualities participants have or need to interact with generative AI \\ 
Challenges            & Challenges that participants face while using generative AI for KE tasks \\ 
Bias                  & Participants opinions related to AI safety testing              \\ 
Generative AI interaction opinions & Participants opinion about the use of generative AI for KE tasks                          \\ \hline
\end{tabular}
\end{table*}

\subsection{Ethics}

The ethnographic study was approved by Authors' institution’s Ethical Advisory Committee via the Full Application Form. Informed electronic (using Microsoft Forms\footnote{\url{https://forms.office.com/Pages/DesignPageV2.aspx}} and storing in CSV) consent was given by the participants. Thirty out of the 39 hackathon participants consented to the ethnographic study. Participants who did not consent were not observed. No personal information was collected in the analysis. 
The documentary data were collected as an output of the hackathon. 

The interview study was approved by Authors' institution’s Ethical Advisory Committee via the Minimal Risk Procedure. Informed verbal (audio recordings and chat transcripts) consent was given by the participants. No personal information was used in the analysis. 

\section{Results}

\subsection{RQ1: What are the main challenges experienced by knowledge engineers when using generative AI for KE tasks?}

Testing the use of generative AI for KE tasks was challenging for all participants. Our observations during the hackathon revealed that the main concerns for participants were identifying appropriate datasets to use, prompting LLMs efficiently, and evaluating the LLM outputs. We asked interviewees which of these they considered important.

\textbf{Dataset}. Interviewees (1), (14), (12) emphasised the importance of datasets as a starting point, and groups used prompting as a means to create a dataset (Observer 1). A dataset was used by groups, for example, to design ontologies based on a domain, to extend existing KGs, and to test reasoning tasks and ontology alignment techniques. Without a dataset, they felt they could not continue with prompting and evaluation (14). ``There are many codes online...to interact with language models that wasn't any problem at all. And the evaluation given that you know, you possess a ground truth,...was not too difficult. So, the most challenging case work for us was finding a dataset'' (8). Finding a dataset was challenging for specific group tasks mainly because of the time pressure (14) and fast planning (1) needed to find and set domain knowledge for the task. On the other hand, an interviewee noted that they felt LLMs would be supportive for establishing datasets. ``The dataset is not what concerns me because if anything, I'm pretty impressed with what LLMs potentially could do in terms of extracting, like I was saying, concepts, relationships, maybe even constraints or parameters or universal restrictions or existential restrictions, from unstructured data'' (13).

\textbf{Prompting}. An interviewee believed prompting is challenging if knowledge engineers' lack NLP training and experience, ``I think prompting templates are going to be, generally more difficult for Semantic Web people, because we are not necessarily natural language processing people'' (1). This was also emphasised in the skills that a knowledge engineer needs to work efficiently with LLMs.  Furthermore, many participants noted the time-consuming phase of iteratively testing prompts to receive desirable outcomes (3), (Observer 1). Some suggested the usefulness of possible ``templates for prompting'' \textcolor{color1}{~\cite{alivanistos-et-al-2023, zhang-et-al-2023-llmke}} (Observer 1), (Observer 2), (3), (13). 
In addition, ``once we settled on a specific prompt, we said okay, it seems like it works good enough, the most tricky part of it was to make the language model consistent in its responses'' (3). This was particularly problematic when the process was automated with thousands of prompting iterations.
Furthermore, syntactical errors also break the scripts and prevent automation (3). 
In contrast, an interviewee pointed out that prompting engineering is ``less scientific'' and models ``with billions of parameters are not controllable,..., resulting in an output that you cannot reproduce one hour later, two days later. So, this might be a problem'' (9).

\textbf{Evaluation}. Participants believed that evaluation is challenging mostly due to the lack of benchmarks for specific KE activities, which depends on manual intervention, ``so there are papers doing auto prompting...I'm a little biased towards evaluation simply because in my tasks evaluation pretty much boils down to a manual evaluation, there's no other way to do it...it's not that we have solved the dataset problem, but we can synthesise some datasets, or we can you know, create them artificially somehow. Evaluation still remains a challenge because it has to be done manually'' (5).
In the case of using a specific controlled dataset for KG construction, the accuracy and coverage of the KG can be automatically evaluated (2), (10). However, in tasks such as designing competency questions or an ontology, there is not always a gold standard to compare with, and manual work from an expert is needed to evaluate the output. Nevertheless, exploring the use of holistic LLM benchmarks such as HELM \cite{liang2022holistic} was suggested (5). \textcolor{color1}{Recent progress in developing benchmarks for LLMs in KE reflects the community’s strong emphasis on establishing standardised methods for their evaluation~\cite{petroni-et-al-2021-kilt, kalo-et-al-2024-lm-kbc}.}

\subsection{RQ2. How do knowledge engineers evaluate generative AI output for their practices?}

We further asked interviewees whether current evaluation techniques for KE tasks could be used to evaluate generative AI outputs. Most respondents believed that evaluation mainly depends on the task, but highlighted that current evaluation metrics are not sufficient for many KE activities. Following this, we asked them what a new benchmark or metric would look like.

\textbf{Current evaluation techniques}.
Some interviewees consider that for specific tasks, like using a controlled dataset (13),(14) and designing ontologies with simple taxonomies (12), evaluation metrics like F1 score, precision and recall (8), and comparison with gold standards are sufficient metrics for evaluation. However, others felt that regarding KE task evaluation, ``it's not [only] a problem that concerns only interaction with LLMs, it's a broader problem that needs to be solved'' (9). Many KE tasks, like ontology semantic and reasoning errors, require human evaluation (1), (5) because ``there is no automatic way [to assess]...how well the ontology represents the domain knowledge'' (11), and this is important because ``if it's mistaking...in relations which creates the hierarchy...[that] is much more important than making a mistake on some data.'' (2)

In addition, others highlighted the need to evaluate the additional knowledge, specifically the extra knowledge produced by LLMS (10) and the knowledge is missing from KGs (4),(6),(14). An interviewee (10) highlighted the extra knowledge produced by LLMs based on their training, for example, when we asked the LLMs to create triples from an expression like ``X was born in Y'', we expected ``X-is a-person'', ``Y-is a-country'', and ``X-birthplace-Y, but we also may get ``Y-part of-European Union''. These outputs include knowledge that may be needed to complete a KG, but it requires evaluation. In addition, a way to evaluate the knowledge missed in the LLMs' outputs is needed. An interviewee pointed out that, ``by using these classical evaluation metrics...[we check] what is inside the [input] text and [we miss] what is not in the text'' (4). 

Finally, interviewees emphasised the difficulty in processing LLMs output mainly because LLMs return ``strings'' (10) of text and thry lacked a way to evaluate the ``generative text'' (7). It is common in KE tasks to use a specific format like JSON or the expression of an ontology. Even in the cases where participants asked the LLM to give output in these particular formats, they faced difficulties because, ``the generated string contains inconsistencies, you really need to parse it and get what you want from the string itself, and then you need to convert it into a list, whatever you need'' (10). \textcolor{color1}{Recent advances in LLMs’ structured generation capabilities have sought to address this challenge, as such requirements have emerged across domains. Approaches such as token-level constrained decoding~\cite{koo-et-al-2024, willard-louf-2023-outlines, dong-et-al-2024-xgrammar} and prompt-level output structuring~\cite{instructor} can enable LLMs to produce outputs in specific, predefined formats.} Complicating this is the fact that LLM outputs are not all the same type (4), for example, a KG includes dates, values, images, URLs etc. It is not possible to have a united way to evaluate the different types. Furthermore, another factor in output evaluation is the similarity of the results (4). 
\textcolor{color1}{Current evaluation techniques primarily rely on exact matching rather than similarity, making them less effective at handling variations and leaving LLMs prone to inconsistency.}

\textbf{New evaluation techniques}.
Most of the interviewees noted that current evaluation techniques are not sufficient and suggested possible alternatives. Some focused on the ontology design (2), (5), suggesting the development of a set of ontologies to be used as gold standards (observer 1). Others suggested the creation of toolkits (1) for LLMs to review ontology errors similar to existing ontology toolkits, like OntoClean~\cite{guarino2009overview} and Oops~\cite{poveda2014oops}. However, they acknowledged that this cannot be applied to large-scale KG like Wikidata, and manual work will still be required. Humans could be more efficient in evaluation if they have the information which prompt is related to the results (Observer 1). \textcolor{color1}{Recent advances in KG evaluation demonstrate that integrating LLMs with human-in-the-loop approaches can be effective~\cite{tsaneva-et-al-2025}.}

Another suggestion was to test the consistency of data and semantics in a real scenario, ``for a company using this data and querying this data, and the simplicity of querying this data, and also the performances of these KGs in queries per second. And also the errors and the consistency of the data that is created,...making sure that the semantics of the data stored in the KG is consistent and reliable'' (9).

In addition, some interviewees were more specific, suggesting the use of techniques used in other fields, like fact-checking (10), adversarial algorithms (13), and self-play from reinforcement learning (12). An interviewee suggested using fact-checking techniques to establish that the extra knowledge received is correct and can be used for the KG construction (10). Another interviewee was fascinated by the concept of adversarial algorithms, a technique used to test machine learning by misguiding a model with malicious input. They suggested using this technique to test LLM results (13). In addition, another suggested the use of self-play technique, a reinforcement learning technique for improving agents' performance by playing ``against themselves". They suggested to ``let the model come up with, a whole host of answers, then compare them between each other [and]...get a more refined answer'', but this will require ``more time and computing'' (12).

Two interviewees (3), (6) commented that we should see generative AI as assistants making the processes in KE smoother and faster because, ``we know that as humans, we are lazy to write down some stuff like the competency question, so, if [you have a suggestion of] competency question you decide [faster], actually, which one you want to use'' (6) and the same can apply every step of the way. They suggested that evaluation should focus on human satisfaction, ``so the language model should probably optimise not really the performances, but the satisfaction of the user at least dealing with that'' (3).

One interviewee suggested that a more ``open-minded'' approach is needed in order to advance the field of evaluation. If metrics are, ``novel, it takes longer to review. If it's a straightforward paper with a slight modification in the method and a new row, it's easier to evaluate the merit of the work. That's not necessarily the way to go. So there's a lot of work to be done in evaluation'' (5).

\subsection{RQ3. What skills does a knowledge engineer need to incorporate generative AI into their practice?}

\textbf{Skills possessed and of use}. As it is all collaborative practices, communication within teams was pointed out as an important skill in multi-disciplinary groups such as those at the hackathon (4), (5). To keep the project moving, the ability to explain and listen to ideas for an efficient workflow are required (4). One interviewee talked about the risk of getting lost in the implementation details, stating, ``I kind of knew how to steer people in a direction that would eventually be useful for the final goals'' (5). 

Having skill in building ontologies was found to be useful in adding context to the prompts (10), (6). For example, using the knowledge structure of a dataset, such as Wikidata (6), and information on how to form triples (10): ``adding...context...as part of the prompt actually helped with the result'' (6). In contrast, one interviewee found that the added context did not make a difference so despite having this skill, they did not consider it important (3).

Building ontologies was also found to be useful in defining the overall goals for the task (5), (9): ``I could contextualise the tasks defined by the hackathon. And also, from a more practical point of view, I knew before starting coding, how the final output would have looked like'' (9).

Having prompting skills and understanding how prompts can be composed and revised to achieve a desired outcome was found to be important (10), (12), (8): ``I think I have a good understanding of the prompt components. So, in a prompt, first you should assign a role, a persona to the language models, and then short task description is really useful. And these models shine when you give a few examples in context learning'' (10). In turn, these participants were able to streamline their approach and prompt more efficiently due to their understanding of the different components of prompts. Even just limited experience was found to be useful for forming prompts with one interviewee describing their skill as ``not scientific'' but found this helped in ``understanding how these models respond to small changes'' (12). Interviewees acknowledged that rules are needed to prompt efficiently. There is a sweet spot in how much information is given to LLMs in order to get a desirable output, with too much information being as problematic as too little.

Coding was the main technical skill that many interviewees mentioned as being important (2), (3), (8), (9). It was found to be useful in interacting with the LLM through the respective API (Application Programming Interface) and not the interface, which was important in querying large batches of prompts. Along with this, knowledge in using repositories of models like Hugging Face\footnote{\url{https://huggingface.co/}} and PyTorch\footnote{\url{https://pytorch.org/}} was also noted as useful by an interviewee (2). A useful skill was in version control systems like Git\footnote{\url{https://git-scm.com/}} to be able to clone, test and debug projects efficiently (2). Some interviewees also mentioned how their coding skill helped them with datasets, such as, scraping websites for datasets (3) and ``[building] a dataset'' (4). 

Developing a scientific framework was found from one interviewee to help define the task's workflow, describing it as ``defining objective, defining hypotheses, defining an evaluation strategy, designing so an experiment'' (7). 

\textbf{Skills gap}. Many of the interviewees who lacked the skill in ontology design and engineering stated its importance (2), (12), (13) specifically for the evaluation of the outputs of the LLM (1), (7). They felt that without this, it is difficult to know what a desired output would look like. 

For those who did not have prior prompting experience, there were two main distinctions on how they viewed this skill. The first way was that some interviewees did not see prompting as a skill and felt that it was too simple of a task with multiple participants, restricting it to a trial-and-error exercise, not acknowledging the possibility to go beyond this. They deemed prompting as requiring no expertise where KE and machine learning were categorised as more skillful (7), (13). One interviewee who described themselves as being familiar with LLMs did not deem prompting as a specialist skill in the KE process as they believed any developer could do it (11). 

On the other hand some interviewees were interested in gaining more experience to go beyond trial-and-error (4), (3), (1), (14). One interviewee acknowledged that prompting can be an intricate process. Some interviewees felt that time for experimentation with generative AI is important to build an understanding of prompting and this familiarity with LLMs can improve performance (5). This can also allow for quicker execution of prompting to streamline the workflow, one interviewee noting, ``I'm pretty sure that a lot of tricks or tips that I don't know about'' (6). 

Various technical skills were mentioned by several interviewees all of which were specific to running LLMs more efficiently. For example, one interviewee mentioned learning to use computer clusters to fine-tune LLMs (10), which needs experience and very high computational requirements. Another interviewee mentioned using LLaMA and Hugging Face for the use of APIs with different models as an important skill (12). There was also an interviewee who felt they needed an enhancement of their coding skills and hardware skills, such as using GPU (9).

\subsection{RQ3. How aware are knowledge engineers of using generative AI responsibly?}

Our question on responsible AI was initially framed in terms of ``safety'', offering the respondents the opportunity to engage with any aspect of AI safety they felt relevant. However, although two respondents raised data security (1) and insufficient data (13) as safety concerns, the majority were not clear on what concerns they might be considering. In these cases, we suggested they focus on bias as most people are familiar with this concept. However, this familiarity may also breed contempt. The authors of this paper observed, ``when asked about harms, they said it wasn’t discussed – assumes humans are biased anyway so there will be bias anyway, to me sounds like implying that there is no need to consider bias'' (Observer 2). One respondent noted that they felt it would be ``interesting and helpful'' to apply safety testing to LLMs and so far had read papers on it (9).

When asked to consider bias, interviewees made comments such as that they had only fleetingly interacted with the issues. ``I’m familiar with the terms, but I haven’t worked on it directly'' (10). They described themselves as have experienced limited exposure (2), possessing a lack of knowledge (12) and interpreting safety as a risk to researchers rather than other publics (14). One respondent had previously worked as an engineer on a bias detection algorithm but felt this was purely a ``theoretical thing''. 

One reason for this lack of engagement appeared to be the perception of safety was seen as a siloed activity rather than integrated with engineering: ``that’s not my field'' (4). This was not necessarily due to a lack of interest, but perhaps also to a lack of opportunity: that participants had been interested in safety work but ``haven’t been able to get to it'' (5). 

The many variants of bias mean that this is a number of problems not one (6). There was some awareness that understanding the type of bias is key to addressing it (4). In terms of specific types of bias, a few interviewees had awareness of what these might be, such as provenance (13), selectivity (5), and gender (6). These had also been brought up with one group in the hackathon, with the observer noting, ``organiser [told the group] – make sure you don’t bias against gender, race, geographical concepts – if competency questions are biased then ontology is biased'' (Observer 2). It is difficult to tell, therefore, whether this knowledge was something the interviewees had applied to previous work, or gained in the relatively recent past of the hackathon.

Correspondingly, there was variation in awareness of bias mitigation: ``[I have] no idea how many safety checking solutions have already been provided'' (1); ``I have some doubts on how these testings are carried out'' (9). One interviewee felt that bias mitigation was more important in commercial products than in the lab (12). There was awareness that addressing bias may change relationships in a KG (4). 

Particular issues related to LLMs regarding provenance and source were noted, for instance, ''there are so many confounding issues in the training of an LLM that even if you address bias in a prompt you can’t see the resources it is using to make a decision'' (6). 
This respondent suggested that there could be ways to mitigate this by mapping the results of repeated prompting. One noted that harm can arise from use, even in a well-tested model. ``So even though [LLMs] pass [bias] tests, I think there's still risk if you prompt them in a hateful or discriminative way'' (10). 

One interviewee suggested that, given the difficulties of mitigating bias, we should at a minimum ''put an array [of] practices in place to monitor what's coming out [and] how unbiased, how ethical, it is'' (5). The idea of 'co-creating' bias mitigation with users was suggested, with one interviewee suggesting not only reviewing the model for bias before use but also asking users for feedback on perceived bias as they used it (7) and another suggesting that, ''either we fix the data, or maybe we fix the tools somehow that [users] can detect this bias and root it out during their working'' (5).

Respondents concerns about difficulties with addressing bias included a lack of knowledge of LLM training data and processes (13), a fear that tests for bias would be ineffective (9), and concern that tests alone are insufficient if appropriate action is not subsequently taken (10). There was also concern about a lack of consensus around safety issues (13) and a feeling that other problems in the model (e.g. accuracy) may be more pressing (2).
 
Some interviewees felt there might be a tension between attempting to reduce bias and model efficacy (2), (12) ''There are evidences that if you debias a language model, it usually downgrades in performance'' (8). Whereas most respondents felt bias was found in data, a few perceived bias might also lie in both the LLM (7) and KGs, specifically in the semantics (9). One respondent felt that safety was not a concern in KGs (14). 
Ever since we conducted the interviews, work has surveyed biases and fairness in LLMs \cite{gallegos-etal-2024-bias}, specifically building taxonomies of papers that propose metrics, datasets, and methods for LLM-based bias evaluation; but without considering LLMs at the task of KG construction.
Since 2024, the Semantic Web community \cite{penuela2024semantic} has studied LLMs for KG construction tasks, e.g. ontology alignment 
\cite{amini2024complexontologyalignmentusing,giglou2024llms4ommatchingontologieslarge}, class membership \cite{allen2024evaluatingclassmembershiprelations}, and conversational ontology engineering \cite{zhang2024ontochat}; although these works recognise the importance of bias and fairness in using LLMs for KG, none of them address them directly.

\subsection{RQ5. What factors may affect knowledge engineers' trust and uptake of the generative AI technology?}

Interviewees have varied opinions on the use of generative AI in their practices. Some find their use promising, and others are more \textcolor{color1}{sceptical}.

\textbf{Promising}.
Interviewees believe that generative AI cannot replace humans (3), but they can support KE activities. Using LLMs we can reduce time spent on tasks (11), (hackathon report), improve communication between experts in different disciplines (hackathon report), perform tasks even without extended experience (12), for example, in ontology design, and incorporate NLP pipelines in our practices without the need to develop them from the beginning (7) \textcolor{color1}{\cite{he-et-al-2024-deeponto, caufield-et-al-2024-ontogpt}}. However, ``we need to use them carefully by providing human oversight, [or] human in the loop, or our pipelines need to add options [where] human can intervene... to avoid passing those errors produced by LLMs to the downstream tasks'' (1).

In this vein, interviewees pointed out that LLMs could improve KE practices (9),(7), (13). However, their use is task-specific and interviewees raised concerns that effective prompting requires experience and there are reproducibility concerns (hackathon report). As well as assisting with processes, there was a suggestion for the incorporation of LLMs to improve existing KE tools (9) or to create new and advanced tools to support KE (13).

\textbf{Sceptic}.
Some interviewees were found to be more sceptical about the use of generative AI for KE tasks. They felt that LLMs \textcolor{color1}{without using any search or retrieval-agumented generation} are not up-to-date with the latest knowledge and that hallucination is very common (1). Moreover, LLMs respond with a prediction of what they think is the answer without giving information about how ``confident'' they are in the answer (13). \textcolor{color1}{Recent work has sought to address this through approaches such as confidence elicitation~\cite{xiong-et-al-2024-uncertainty, geng-et-al-2024}, self-assessment~\cite{ren-et-al-2023-self-evaluation}, chain-of-verification~\cite{dhuliawala-et-al-2024-cove}, and repeated sampling~\cite{brown-et-al-2024}.}
In reverse, trusted KGs already support AI with trusted knowledge and can also support LLMs (1). 

Other interviewees highlighted that LLMs cannot perform all KE activities (11), and the lack of evaluation techniques (6) lead some of the participants to find no value in their use. An interviewee expressed, `` Yes, it can help you to quickly get some data just for testing purposes, for example. But for production, I don't think that I will use it...since I never know which part of it is actually correct'' (6).
During the hackathon, participants felt that LLMs can perform some tasks or simple examples, but in complex scenarios, the output cannot be controlled (i.e. evaluation). An interviewee expressed that integrating KGs and LLMs was not required as LLMs would entirely replace KGs. 

\section{Discussion}

In this section, we consider some of the key issues that arose from our results. 
We examine in more depth issues related to skills, in particular prompting and bias detection and mitigation, and conclude by emphasising the importance of transparency and explainability in KGs. Finally, we introduce the concept of ``KG Cards'' and we suggest the use of the existing Model Cards for the KG embedding models.

\subsection{Issues in Using Generative AI for Knowledge Engineering}

\textcolor{color1}{A fundamental challenge in using generative AI for KE stems from the ambiguity and variability of its outputs. This is most evident in the inherent irreproducibility of results from single, directly prompted LLMs, where generated relations or term similarities are based on statistical probability rather than consistent fact. Such inconsistency undermines scientific rigour and complicates error tracing, making control and refinement difficult. \textcolor{color1}{(Since we conducted the interviews, this has been at least partly mitigated with retrieval agumented generation (RAG) \cite{lewis-et-al-2020-rag}, which enables LLMs to retrieve relevant documents and integrate their information to user prompts before providing query responses; however, issues of bias and fairness around the integration of specific external knowledge sources and the models themselves persist \cite{kim-etal-2025-mitigating,wu-etal-2025-rag}.}) These concerns align with prior findings that knowledge extraction remains challenging despite advances in NLP~\cite{allen2023knowledge,noy2019industry}. Compared to pre-LLM human-in-the-loop KE tools~\cite{ke-pre-llm}, LLM-assisted approaches offer more end-to-end workflows and streamlined human–machine interaction, but also obscure the origin of errors.
However, recent advances---such as tool calling~\cite{qin-et-al-2023-toolllm}, retrieval-augmented generation~\cite{lewis-et-al-2020-rag,gao-et-al-2023-rag}, confidence elicitation~\cite{xiong-et-al-2024-uncertainty, geng-et-al-2024}, and self-assessment~\cite{ren-et-al-2023-self-evaluation}---have significantly mitigated these issues. These developments could lead to more reliable, and reproducible LLM-assisted KE workflows, thus change the attitude of knowledge engineers towards LLM-assisted systems.} 

Another notable application of LLMs in knowledge engineering is their potential to generate structured datasets from various unstructured data sources, particularly for data integration purposes. Research has shown that, when given a prompt containing the term ``dataset'', in the absence of a dataset, ChatGPT may attempt to ``curate'' a dataset like an approximation from unstructured data~\cite{walker2023prompting}. Therefore, this application seems highly plausible for KE and, particularly, knowledge extraction in a structured format which is familiar to existing practices. Following the usual actions of cleaning and transforming, the dataset can be used for KG construction.\textcolor{color1}{Recent advances in the structured generation capabilities of LLMs have further strengthened their potential for such applications~\cite{koo-et-al-2024, willard-louf-2023-outlines, dong-et-al-2024-xgrammar, instructor}.} 
Yet, the accuracy of LLM results remains an issue.


Following the knowledge extraction and integration, another particular area of concern was evaluation. Attempting to evaluate LLM support for KE tasks would simply exacerbate existing evaluation challenges, which were many. These include evaluating \textit{accuracy} (i.e., syntactic and semantic) and \textit{coherency} (i.e. consistency and validity), emphasising that we may assess accuracy automatically, but most of the time, coherency requires manual interventions. Furthermore, concerns about \textit{coverage} (i.e. completeness and representativeness ) and \textit{succinctness} (i.e. conciseness and understandability) are also relevant to input as well as output, and should be included as a consideration. This is a similar problem to that experienced in crowd-sourced KGs such as Wikidata \cite{abian2017wikidata}, where variations in the input (such as background and skill of contributors) also demands evaluation as much as the output.
\textcolor{color1}{Recently, the community has devoted increased attention to evaluation, emphasising both the development of robust evaluation benchmarks and the adoption of human-centered principles, strategies, and paradigms~\cite{zhang-et-al-2025-evaluation,llugiqi-et-al-2025,lippolis-et-al-2025,heim-et-al-2025,hannah-et-al-2025}.} Recent research has explored the use of LLMs as a judge to evaluate the results of competency questions, but suggests these are still at the stage of being useful as supporting human inputs, rather than as trustworthy standalone tools \cite{kommineni2024humanexpertsmachinesllm}.
Regarding the evaluation of generative AI outputs, one interesting suggestion was made related to using adversarial algorithms. ``Would something or a product like that stand the scrutiny of an adversarial network, who is trying to maybe test or break the new information derived by the existing system. So can there be a counter point counter punch that tries to break it, and if it cannot be broken or falsified?'' An adversarial attack on BARD~\cite{BARD2024} asking to repeat one iteratively one word eventually got the model to reveal gigabits of its training data, demonstrating the usefulness of such processes in testing systems~\cite{nasr2023scalable}.

Another proposed aspect of evaluation was human satisfaction. Other research has shown that the natural language element of LLMs, often read by humans as possessed of emotion, can make people feel that the process of using the model was successful even if the end result was not \cite{walker2023prompting}. This could be potentially problematic for KE tasks, which requires logic over feelings. A comparable, well-researched situation is that of emotion in financial markets, which can move prices up or down based on public mood rather than underlying economic fundamentals.

\subsection{Key Skills for Effective Prompting}

Prompting was a fascinating area of discussion. Although many interviewees spoke of how they found previous skills useful in developing effective prompts (and equally, that it was very possible to use ineffective prompts by accident), some interviewees also felt that prompting skills were not important. This conflict in opinions may be explained by the different tasks and the experience levels of interviewees in KE and in research. 

However, ultimately, prompting was utilised for all the different KE tasks. This flags a new era and requirement for a new skill set in the field of KE. This suggests that the creation of templates \textcolor{color1}{(e.g., \cite{alivanistos-et-al-2023, liu2023pre, zhang-et-al-2023-llmke})} for the KE tasks could benefit the field towards KG construction and fill in gaps in prompting skill level. This has parallels in the introduction of crowd-sourcing to KG construction around a decade ago. Using crowdsourcing means for large-scale KG construction was a successful innovation supporting many intelligent applications today~\cite{acosta-et-al-2013-crowdsource, revenko-et-al-2018-crowdsource, kou-et-al-2022-crowdgraph}. Furthermore, leveraging (semi-)automatic prompting methods~\cite{ye2023prompt} and incorporating domain-specific prompting strategies~\cite{prompt-engineering-medical-education} can help address skill gaps and improve performance. Very recent research suggests that although reasoning has received a lot of attention as a prompting strategy, for some major LLMs this does not provide substantial improvement over simple instruction and task demonstration for knowledge extraction from text \cite{doi:10.3233/SW-243719}. This implies that in this area at least, proficiency in prompting may easily be acquired. 

\subsection{Key Skills to Mitigate Bias}

Knowledge engineers and domain experts involved in generative AI-supported KE projects are key stakeholders in promoting responsible AI practices. Ensuring that AI systems are developed, assessed, and deployed in a safe, trustworthy, and ethical manner is critical for fostering appropriate trust among these professionals. There was a distinct lack of engagement with the processes of bias mitigation. This tended to arise from limited exposure to techniques or theories for identifying bias in the model.  While this was sometimes due to a lack of opportunity for knowledge engineers to acquire capabilities, in other cases it was due to a lack of a sense of responsibility for bias mitigation. It is difficult to argue that the identification and eradication of bias should be left to a specific set of experts, rather than integrating core safety facets into the training of those responsible for building AI. Given the variety of safety and ethical aspects and potential harms in AI, it may be that space for ``ethical deliberation'' in the KE process is more valuable than simply providing, for instance, a list of biases for engineers to look out for~\cite{gogoll2021ethics}.

\textcolor{color3}{While capacity building (training) and bringing awareness to the importance of taking responsibility for ownership of AI safety issues appear to be two separate issues, they are contingent upon each other and should be treated as two sides of the same coin. Proper exposure to bias awareness and mitigation techniques should include training on when and where to review aspects of safety – thus integrating the idea of responsibility for safety with knowledge of the actions to take. }

a

Our results illustrated a data-centric approach to bias in KE, where our participants felt that the source of bias was more likely to be the data than the graph model itself, and relatively few recognised that the model may be a source of bias.  Including generative AI in the construction of KGs will introduce a further source of bias  \cite{huang2023bias}.  The integration of generative AI into the KE processes may additionally create further potential for harm, such as when, intentionally or unintentionally, the model can be manipulated to overcome privacy (as shown in the adversarial attack example above \cite{nasr2023scalable}). This suggests an integrated review of the possible sources of harms introduced to KE when generative AI models are utilised would be valuable.

\textcolor{color1}{A number of the issues raised by participants in the hackathon  - particularly those around evaluation and ethics – speak to deeper, more sustained challenges within the discipline.} 
\textcolor{color1}{Our findings show that many knowledge engineers feel ethical issues are the responsibility of  'someone else, and that bias mainly resides in data and not in any part of model construction. This finding is similar to other studies with AI engineers or practitioners, who feel either that the responsibility for mitigating harm does not lie with them,  or that as individuals they cannot do much about it \cite{10.1145/3632121}. A great deal of recent work on LLMs for KE emphasises that research into bias in this area is of great importance  \cite{gallegos-etal-2024-bias,penuela2024semantic,amini2024complexontologyalignmentusing,giglou2024llms4ommatchingontologieslarge,zhang2024ontochat}. However,  no study has directly addressed bias and fairness in LLM-based KG construction and there are no concrete proposals on how to address it.  This suggests that work needs to be done to embed a greater understanding of where bias lies, and how KE can address it.  LLMs are not neutral and have bias up and down the chain that must be understood and appraised \cite{kim-etal-2025-mitigating,wu-etal-2025-rag}. Benchmarks for evaluating LLM-generated ontologies have been proposed \cite{EURECOM+7945}, but are limited to assessing the presence/absence of classes and properties and their level of abstraction, not their possible issues around bias and fairness. A starting point would be measuring bias and fairness in LLM-based KG outputs, for which inspiration could be taken from concepts such as Vendi (diversity evaluation) scores and the multii-metric HELM approach \cite{friedman2023vendiscorediversityevaluation,liang2022holistic}.}

\subsection{KG Cards to Enhance Transparency}

As described in Section \ref{sec:ethics}, prior research aimed at improving transparency and explainability has introduced frameworks such as Model Cards \cite{mitchell2019model}, Data Cards \cite{pushkarna2022data}, and Data Statements \cite{bender2018data}. Implementing our review of harms we point out the need for similar practices for KGs. The aforementioned Cards can not fully serve the KG's needs because they do not consider the unique design aspects of KGs. These Cards fail to describe the ontology and the KG data, such as relations and schemas. 
Building on these studies, we propose the concept of Knowledge Graph Cards (\textbf{KG Cards}) as a documentation framework to promote transparency and explainability, ensuring the creation and use of trustworthy KGs. Drawing inspiration from the flexibility and extensibility of Data Cards, and considering the foundational representation format of most KGs, KG Cards are designed to enhance transparency and trustworthiness by providing comprehensive and structured documentation.

\begin{figure*}[h]
  \centering
  \includegraphics[width=0.88\linewidth]{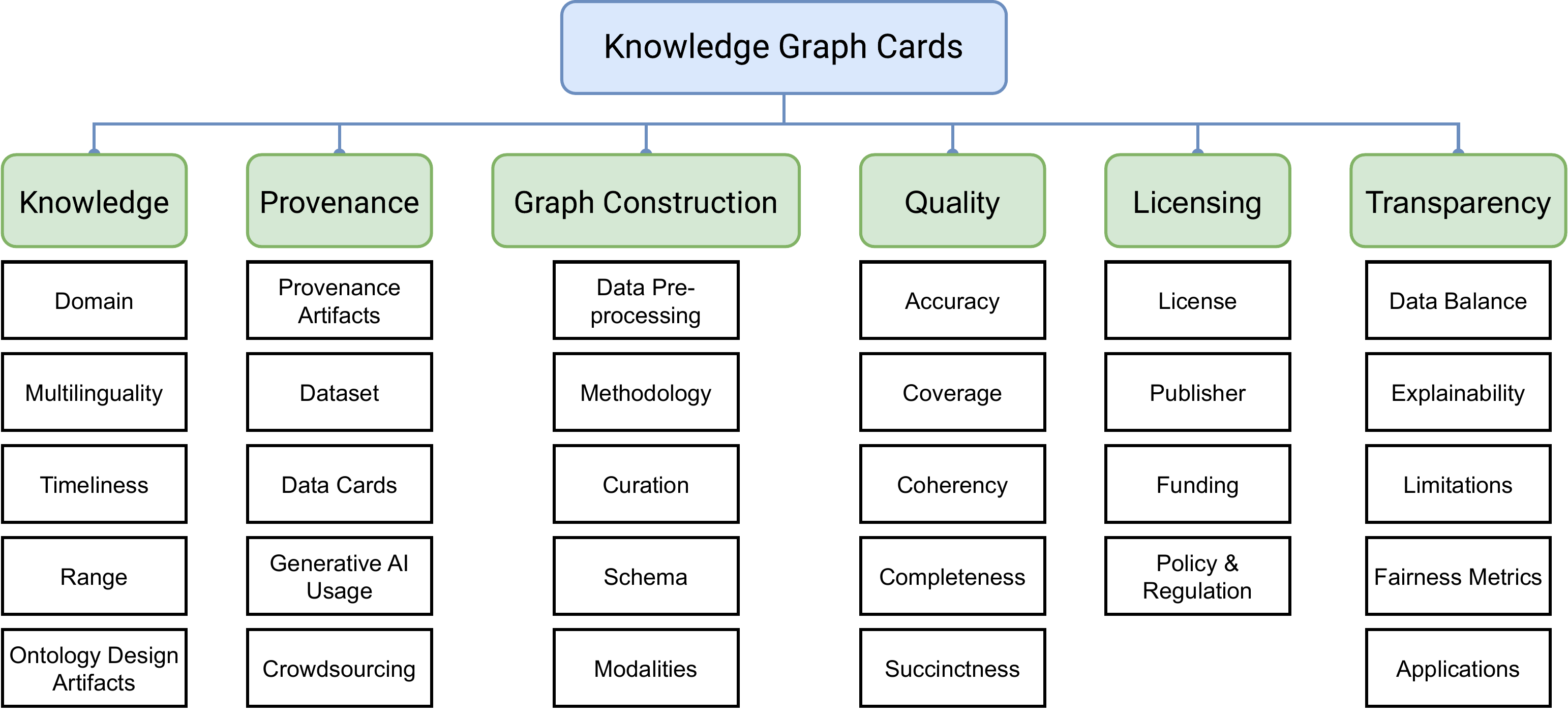}
  \caption{Six suggested sections of KG Cards with related details.}
  \label{fig:kg-card}
\end{figure*}

As shown in Figure~\ref{fig:kg-card}. It contains 6 main themes: 
\begin{itemize}
    \item Knowledge. The Knowledge section aims to provide an overview of the domain, the languages represented, and the range of knowledge included. It may also document ontology development artifacts, such as competency questions, which help define the knowledge graph’s (KG) scope and support ontology testing.
    \item Provenance. The Provenance section is intended to capture detailed provenance information about the KG. Various platforms and frameworks, such as the W3C Provenance Ontology (PROV-O)~\cite{prov-o}, have been proposed for managing provenance in KGs~\cite{kleinsteuber2024managing}. Additional provenance details can be documented here, including information about crowdsourcing (e.g., platform used, demographics of crowdworkers, and domain experts), references to datasets (and Data Cards, where applicable) used in KG creation, and other contextual information.
    \item Graph Construction. The Graph Construction section outlines the details of the data preprocessing methods and the approaches used during the construction and curation process. This includes techniques for knowledge extraction, integration, and the use of KG embedding models.
    \item Quality. A crucial aspect of KG transparency and explainability comprises the quality assessment. Considering the KG quality assessment dimensions described in Section~\ref{sec:ke} and quality dimensions for linked data~\cite{zaveri2016quality}, we introduce the Quality section. This includes descriptions about accuracy, coverage, coherency, and succinctness. Many studies have focused on completeness \cite{heist2020knowledge,Paulheim2017KnowledgeGR,seo2022structural,issa2021knowledge}, which is also a vital dimension for explainability. Syntactic and semantic accuracy describes how errors are handled, while timeliness states the pace of updates. Coverage looks at the network details, stating statistics which can offer an understanding of the KG coverage and potentially can support the next card about safety tests. Moreover, coherency expresses contradictions in the data and succinctness describes the human-readable aspects, the URIs details, SPARQL query services, and community contact options.
    \item Licensing. The Licensing section provides details about the licenses, publishers, funding sources, and associated policies.
    \item Transparency. While previous documentation methods provide detailed transparency principles, they often lack emphasis on fairness aspects. ``Fairness'' as a concept is vague, culturally dependent and difficult to operationalise \cite{munn2023uselessness,Dodds_2020}. For KG Cards, we suggest the Transparency Card with a set of transparency and fairness checks related to data, the intended and unintended application of the KG, and explainability. With this, we aim to provide explanations and support for bias and ethics.
\end{itemize}

An example of the use of KG cards is presented in Appendix~\ref{sec:appenD}, Table~\ref{tab:kg-cards}. The presented KG Cards represent a preliminary framework for enhancing transparency in KGs. Like previous model and dataset documentation methods, the proposed KG Cards are adaptable, allowing more sections and properties to be included based on the specific needs of the stakeholders in various contexts. Further research could offer valuable insights through extended analysis, helping to validate usability and identify potential modifications or enhancements to the structure.

The use of well-structured documentation for transparency and explainability has inspired initiatives like the Hugging Face platform's adoption of Dataset Cards and the Croissant metadata format for its shared datasets~\cite{croissant}.\footnote{\url{https://huggingface.co/docs/hub/datasets-cards}} 
Table~\ref{tab:alignment} shows the alignment between Hugging Face's Dataset Cards, Croissant metadata format, and our suggested KG Cards. Most cards match different needs and directions depending on the product (i.e., dataset or KG); however, we suggest incorporating state-of-the-art metrics and emphasising the importance of ethical issues.

\begin{table*}[h]
\caption{Alignment between KG Cards, Hugging Face dataset cards, and Croissant metadata format.}
\label{tab:alignment}
\begin{tabular}{ccc}
\toprule
\textbf{KG Cards} & \textbf{Hugging Face Dataset Cards} & \textbf{Croissant Metadata Format} \\ \midrule
Knowledge         & Dataset Description                 & The Dataset Metadata Layer  \\
Provenance        & Dataset Creation                    & The Resources Layer         \\
Construction      & Dataset Creation                    & The Structure Layer         \\
Quality           & Dataset Structure                   & The Dataset Metadata Layer  \\
Licensing         & Additional Information              & The Dataset Metadata Layer  \\ 
Transparency      & Considerations of Using the Data    & The Dataset Metadata Layer (RAI properties), The Semantic Layer \\ \hline
\end{tabular}
\end{table*}

\subsection{Model Cards to Reduce Harms}

Besides the Data Cards framework, Hugging Face has also adapted the set of Model Cards for transparency and detailed documentation of their models.\footnote{\url{https://huggingface.co/docs/hub/model-cards}} We suggest a similar tactic for the explainability of KG embedding models in the KE field. The set of Model Cards~\cite{mitchell2019model} can be used to support documentation with a metadata template and enhance model transparency. Table~\ref{tab:KG-embedding-MC} presents the list of Model Cards with descriptions and one example of the KG embedding model, TransE~\cite{bordes-et-al-2013-transe}. TransE is a KG embedding model that models relationships in the graph by interpreting them as translations applied to the low-dimensional embeddings of entities. Inspired by the TransE model, several variant models have been developed, such as TransR (rotating)~\cite{lin-et-al-2015-transr} and TransF (folding)~\cite{feng2016knowledge}. The TransE example in the table shows that KG embedding model studies offer detailed documentation for model training and evaluation, but the lack of demographic details and ethical considerations flag the need, similar to datasets, to adapt specialised instructions for model documentation in the field of KE.

\begin{table*}[h]
\caption{Model Cards by Michelle et al. \cite{mitchell2019model}, with descriptions and the alignment for one example of TransE model.}
\label{tab:KG-embedding-MC}
\begin{tabularx}{\textwidth}{>{\hsize=.6\hsize}X>{\hsize=.7\hsize}X>{\hsize=.7\hsize}X}
\toprule
\makecell[c]{\textbf{Model Cards}} &
  \makecell[c]{\textbf{Description}} &
  \makecell[c]{\textbf{Example Model (TransE)}} \\\midrule
Model details &
  Basic information about the model &
  \makecell[l]{Bordes et al. \cite{bordes2013translating}\\ 2013\\ Translation-based model\\ \{bordesan, nusunier, agarciad\}@utc.fr\\ \{jweston, oksana\}@google.com} \\\hline
Intended use &
  Use cases that were envisioned during development &
  Link prediction (details in \cite{bordes2013translating} section 4.3) \\\hline
Factors &
  Factors could include demographic or phenotypic groups, environmental conditions, technical attributes, or others &
  \makecell[c]{-} \\\hline
Metrics &
  Metrics should be chosen to reflect potential real-world impacts of the model &
  Head and Tail example (details in \cite{bordes2013translating} Table 5) \\\hline
Evaluation data &
  Details on the dataset(s) used for the quantitative analyses in the card &
  WordNet \cite{fellbaum2010wordnet} and Freebase \cite{bollacker2008freebase} (details in \cite{bordes2013translating} section 4.1) \\\hline
Training data &
  May not be possible to provide in practice. When possible, this section should mirror Evaluation Data. If such detail is not possible, minimal allowable information should be provided here, such as details of the distribution over various factors in the training datasets &
  Trained on a large-scale split of Freebase containing 1M entities, 25k relationships and more than 17M training samples (details in \cite{bordes2013translating} section 4.1) \\\hline
Quantitative analysis &
  Unitary and intersectional results &
  Mean and Hits metrics, compared with RESCAL \cite{nickel2011three}, SE \cite{bordes2011learning}, SME\cite{bordes2014semantic}, LFM \cite{jenatton2012latent} (details in \cite{bordes2013translating} section 4.2) \\\hline
Ethical considerations &
  Demonstrate the ethical considerations that went into model development, surfacing ethical challenges and solutions to stakeholders &
  \makecell[c]{-} \\\hline
Caveats and recommendations &
  List additional concerns that were not covered in the previous sections &
  \makecell[c]{-} \\\hline
\end{tabularx}
\end{table*}

\section{Conclusion}

Generative AI heralds the emergence of a new era for many fields. It is inevitable that they will change KE in many ways. In what direction this will lead us is not clear. The hype around conversational generative AI is such that may one day render KGs obsolete. Currently the volatility and unreliability of generative AI means that KGs will continue to be a trusted source, although it is impossible to tell if this will still be true if LLMs become more reliable in the future. Today, however, the added efficiency in terms of addressing volume based tasks is of great potential for use alongside more traditional methods. In terms of adoption of new ways of creating KGs this recent development is somewhat analogous to the introduction of crowd sourcing to KE; an innovation that has subsequently proved of immense value. However, it is crucial that those working in the field are appropriately skilled and supported to work with generative AI in an effective, productive and safe fashion. As well as training, this implies the need for documentation to accompany LLM-assisted KGs such as prompt templates and KG cards. 

\subsection{Limitations} 
Participants were selected from European research labs that have a strong profile in either publishing in KE venues, or in offering well-used KE industry products. Given the number of participants not all KE tasks were investigated, the experiments were relatively small scale and groups worked for only three days on the tasks. Furthermore, most interviewees were different-level PhD students, missing the opinion of researchers with more than 10 years of experience in the field. \textcolor{color1}{It is worth mentioning as a limitation that there is a gap between the completion of data analysis and the publication of this study, which may influence the contemporaneity of the findings. However, the core insights remain robust and continue to provide meaningful contributions to how knowledge engineers use generative AI. }

\subsection{Contributions and Future Work}
This is one of the first attempts to understand the interaction between knowledge engineers and generative AI. We have identified and presented key areas of strength and challenge for KE working with LLMs, and we have built on the suggestion of implementing Data Cards for KGs and propose extending these to more comprehensively address safety within KGs developed with generative AI. 

Future work could explore how generative AI can \textcolor{color1}{solve other KE tasks and assist}less experienced users in KE tasks, building on its established applications in other domains such as programming, report generation, and content creation. Additionally, further efforts are needed to deepen our understanding of the role generative AI plays in specific KE tasks and throughout the KG lifecycle. A particular focus could be placed on conducting comprehensive studies to identify practical strategies for detecting and mitigating biases in LLM-assisted KGs across the entire development pipeline. \textcolor{color3}{Key to this is understanding the scale of whether KE's lack knowledge about safety, or whether the real barrier is obstacles to the implementation of this knowledge. }

\subsection{Acknowledgments}
This paper is produced as part of the MuseIT project which has been co-funded by the EU under the Grant Agreement number 101061441. The authors would like to thank all the participants in the 2023 Knowledge Prompting Hackathon at King's College London. 

\bibliographystyle{cas-model2-names}
\bibliography{cas-refs}

\appendix

\clearpage
\section{Supplementary Materials for Interview Study}
\label{sec:appenA}

\paragraph{Consent} Consent text (to read to interviewees):

With this research, we aim to investigate how knowledge engineers and practitioners use generative AI to develop and maintain KGs, ontologies, and knowledge bases. We will not collect any personal data. The responses to the questions are fully anonymous. The collected data will be shared with Otter AI, an automated transcription service hosted on an AWS server in the US. Your data will be processed under the terms of UK data protection law (including the UK General Data Protection Regulation (UK GDPR) and the Data Protection Act 2018). Transcripts will be kept until the publication of the research. 

If you consent to participate in this study, please say loud and clear ``Yes I consent''.

\paragraph{Introduction}
Give a bit of an intro of what are the main topics of the questions (e.g., your experience during Hackathon, challenges, skills, automation, evaluation, responsible AI). In the end we will ask you about anything important you want to mention, and you will have time to add comments. Mention that there is no wrong answer.

\paragraph{Questions} 
\begin{enumerate}
\item What is your experience with KE and building KGs and ontologies? 
\item How did you find the process of interacting with LLMS to solve KE tasks? 
\item We found that during the Hackathon it was challenging for the groups to (i) find a dataset for testing their pipelines, (ii) find the right prompt questions for the LLMs, and (iii) evaluate the quality of each step toward building an ontology. Which one do you think is the most important? 
\item What do you think were the most important skills you have to complete the tasks during the Hackathon?
\begin{enumerate}
    \item Do you feel you had those skills, or do you need to gain experience to complete several tasks? 
\end{enumerate}
\item Was there any automated process in the project you worked on during the Hackathon?
\begin{enumerate}
    \item What was the purpose of this automation? 
\end{enumerate}
\item Currently, to evaluate ontologies and KG constructions, we use metrics like precision and recall against gold standards, or human evaluation (semantic metric). Do you think this is enough? 
\begin{enumerate}
    \item What they would look like new benchmarks and metrics? 
\end{enumerate}
\item Have you previously used safety testing for responsible AI? Safety testing can be Bias testing against discrimination, Ethical considerations, risk assessment, test for long-term effects.  
\begin{enumerate}
    \item Do you think this can be used in the LLMs KE scenarios or we need to develop specific solutions? 
\end{enumerate}
\item Do you think there is anything else important for us to know related to LLMs and KE? 

\end{enumerate}

\section{Codebook for Thematic Analysis}
\label{sec:appenB}
See Table~\ref{tab:codebook}.

\begin{table*}[b]
\caption{Codebook created by applying inductive and deductive thematic analysis.}
\label{tab:codebook}
\begin{tabularx}{\textwidth}{>{\hsize=.45\hsize}X>{\hsize=.65\hsize}X>{\hsize=.9\hsize}X}
\toprule
\textbf{First level code} & \textbf{Second level code} & \textbf{Third level code}                                                                  \\\midrule
Background Experience &                                                      &                                                                     \\
                      & KG construction              &                                                                     \\
                      & KG explainability            &                                                                     \\
                      & Machine Learning             &                                                                     \\
                      & Ontologies                   &                                                                     \\
                      & Multidisciplinary background &                                                                     \\\hline
Bias                  &                                                      &                                                                     \\
                      & Bias as risk                 &                                                                     \\
                      & Bias mitigation awareness    &                                                                     \\
                      & Difficulty of assessing bias &                                                                     \\
                      & Limited awareness            &                                                                     \\
                      & Other safety issues          &                                                                     \\
                      & Removing bias                &                                                                     \\
                      & Types of bias                &                                                                     \\\hline
Challenges            &                                                      &                                                                     \\
                      & Dataset                      &                                                                     \\
                      & Evaluation                   &                                                                     \\
                      &                                                      & KE tasks still demand manual evaluation     \\
                      &                                                      & Lack in KE evaluation techniques            \\
                      & Prompting                                            &                                                                     \\\hline
Evaluation            &                                                      &                                                                     \\
                      & Current evaluation techniques                        &                                                                     \\
                      &                                                      & Evaluate the new knowledge produced by LLMs \\
                      &                                                      & KE evaluation techniques are not sufficient \\
                      &                                                      & LLMs output is hard to process              \\
                      & New evaluation techniques                            &                                                                     \\\hline
LLM use opinions      &                                                      &                                                                     \\
                      & Promising                                            &                                                                     \\
                      &                                                      & LLMs can support humans                     \\
                      &                                                      & LLMs could improve KE tools                 \\
                      & Skeptic                                              &                                                                     \\
                      &                                                      & LLMs are not trusted                        \\
                          &                            & \makecell[l]{LLMs cannot perform all KE tasks and still need \\ human-in-the-loop} \\\hline
Skills                &                                                      &                                                                     \\
                      & Skills had and helped                                &                                                                     \\
                      &                                                      & Communication                               \\
                      &                                                      & Knowledge or ontology engineering           \\
                      &                                                      & LLM                                         \\
                      &                                                      & Technical skills                            \\
                      & Skills need to have                                  &                                                                     \\
                      &                                                      & LLM                                         \\
                      &                                                      & Ontology design or ontology engineering     \\
                      &                                                      & Technical skills               \\\hline          
\end{tabularx}
\end{table*}

\section{List of Tools and Methodologies}
\label{sec:appenC}
See Table~\ref{tab:tools}.

\begin{table*}[h!]
\caption{KE and generative AI tools and methodologies with description and references.}
\label{tab:tools}
\begin{tabularx}{\textwidth}{>{\hsize=.55\hsize}X>{\hsize=1.45\hsize}X}
\multicolumn{2}{c}{(a) Knowledge Engineering} \\
\toprule
\textbf{Tools \& Methodologies} & \makecell[c]{\textbf{Description}} \\ \midrule
eXtremeDesign methodology~\cite{blomqvist2016engineering}   & A framework for pattern based ontology design          \\ \hline
NeOn methodology~\cite{suarez2011neon}            & A scenario-based methodology that supports the collaborative and dynamic aspects of ontology development, including in distributed environments       \\ \hline
HermiT Reasoner~\cite{glimm2014hermit}             & A reasoner for ontologies that, given an OWL file, can determine if the ontology consistent, identify subsumption relationships between classes, and more     \\ \hline
OOPS~\cite{poveda2012validating}                      & A web-based tool, independent of the ontology development environment, used for detecting modelling errors     \\ \hline
OAEI~\cite{euzenat2011ontology}                       & Aims at comparing ontology matching systems on precisely defined test cases which can be based on ontologies of different levels of complexity and use different evaluation modalities       \\ \hline
OntoClean~\cite{guarino2002evaluating}                 & A methodology for validating the ontological adequacy of taxonomic relationships based on formal, domain-independent properties of classes          \\ \hline 
& \\
\multicolumn{2}{c}{(b) Generative AI} \\ \toprule
\textbf{Tools \& Methodologies} & \makecell[c]{\textbf{Description}} \\ \midrule
ChatGPT~\cite{ChatGPT}                      & A language model developed by OpenAI         \\ \hline
PaLM~\cite{anil2023palm}                    & A transformer-based model developed by Google    \\ \hline
Llama~\cite{touvron2023llama}               & A family of transformer-based autoregressive causal language models developed by Meta       \\ \hline
mPLUG-Owl~\cite{ye2023mplug}                & A multimodal model that integrates images, text, and other information for comprehension and to generate responses     \\ \hline
InstructBLIP~\cite{dai2024instructblip}     & A vision-language model leveraging instruction tuning to achieve state-of-the-art performance    \\ \hline
Claude~\cite{AnthropicClaude}               & A language model developed by Anthropic with use cases such as summarisation, Q\&A, and coding     \\ \hline
\end{tabularx}
\end{table*}

\section{Example of KG Cards}
\label{sec:appenD}
See Table~\ref{tab:kg-cards}.

\begin{table*}[h!]
\caption{KG Cards for the CS-KG~\cite{dessi2022cs}. The CS-KG is accessed through \url{https://scholkg.kmi.open.ac.uk/}. The dumps include the updated versions of the KG. The methods used for the KG's development, the data, and some statistics are presented in the CS-KG scientific publication~\cite{dessi2022cs}. The publication describes the first version of the KG.}
\label{tab:kg-cards}
\begin{tabularx}{\textwidth}{|>{\hsize=.32\hsize}X|>{\hsize=1.68\hsize}X|}
\multicolumn{2}{l}{\textbf{(a) \textit{Knowledge}}} \\ \hline
\makecell[c]{\textbf{Card}}      & \makecell[c]{\textbf{Description}}     \\ \hline
Domain             & Computer Science \\ \hline
Multilinguality     & English         \\ \hline
Timeliness        & The CS-KG includes statements and entities extracted from scientific papers in the period 2010-2021. The team plans to keep maintaining and updating CS-KG in the following several years. The team created a fully automatic pipeline that we will run every six months to produce new versions of CS-KG that will include recent papers from OpenAlex~\cite{dessi2022cs}.             \\ \hline
Range & 3.9M entities are classified as \textit{Methods}, 1.3M as \textit{Tasks}, 450M as \textit{Materials}, 215K as \textit{Metrics}, and 4M are associated with the \textit{OtherEntity}. \\ \hline
Ontology Design Artifacts & The CS-KG ontology builds on top of SKOS (https://www.w3.org/2004/02/skos/) and PROV-O (https://www.w3.org/TR/prov-o/). The design of the object properties started from a set of 39 high level predicates (e.g., uses, analyzes, includes) produced by the knowledge graph generation pipeline (see Section 4.2). The team then associate specific domain and range constraints to them, which are used to drive and correct the automatic extraction process~\cite{dessi2022cs}. \\ \hline
\end{tabularx}
\end{table*}

\begin{table*}[h!]
\begin{tabularx}{\textwidth}{|>{\hsize=.32\hsize}X|>{\hsize=1.68\hsize}X|}
\multicolumn{2}{l}{\textbf{(b) \textit{Provenance}}} \\ \hline
\makecell[c]{\textbf{Card}}      & \makecell[c]{\textbf{Description}}     \\ \hline
Provenance Artifacts & PROV-O. Each statement in CS-KG includes: \texttt{provo:wasDerivedFrom}, which provides provenance information and lists the MAG IDs (now OpenAlex IDs) of the articles from which the statement was extracted; \texttt{provo:wasGeneratedBy}, which provides provenance and versioning information of the tools used to detect the statement~\cite{dessi2022cs}. \\ \hline
Dataset \& Data Cards         &   Scientific articles selected by considering all papers from 2010 to 2019 with at least 1 citation (as of December 2021) and all the papers in 2020-2021 period from the set of articles from MAG~\cite{mag} associated with the Field of Study ``Computer Scienc''. Since MAG has been decommissioned in 2021, the following versions will adopt OpenAlex (https://openalex.org/), which offers a comparable publication coverage~\cite{dessi2022cs}.        \\ \hline
Generative AI Usage & No information \\ \hline
Crowdsourcing       & No use of crowdsourcing. 
\\ \hline
\end{tabularx}
\end{table*}

\begin{table*}[h!]
\begin{tabularx}{\textwidth}{|>{\hsize=.32\hsize}X|>{\hsize=1.68\hsize}X|}
\multicolumn{2}{l}{\textbf{(c) \textit{Graph Construction}}} \\ \hline
\makecell[c]{\textbf{Card}}      & \makecell[c]{\textbf{Description}}     \\ \hline
Data preprocessing & 
  No information      \\ \hline
Methodology           &   
  Based on the website (\url{https://scholkg.kmi.open.ac.uk/}), the KG was generated by applying an automatic pipeline that extracts entities and relationships using four tools: DyGIE++, Stanford CoreNLP, the CSO Classifier, and a new PoS Tagger. CS-KG publication \cite{dessi2022cs} presents the full pipeline.

  This pipeline was evaluated on a manually crafted gold standard yielding competitive results. It then integrates and filters the resulting triples using a combination of deep learning and semantic technologies in order to produce a high quality KG. The CS-KG publication \cite{dessi2022cs} states that 1200 statements were selected for evaluation. The set was manually annotated by 3 senior computer science researchers. The Fleiss’ kappa agreement \cite{fleiss1979large} between the annotators was 0.435, indicating a moderate agreement. The majority vote schema was employed to generate the gold standard. In order to show the advantage of the hybrid method that builds on top of multiple tools, the authors compared the CS-KG pipeline against DyGIEpp, OpenIE, PoST, and against the union of their results (DyGIEpp + OpenIE + PoST). Table 2 in \cite{dessi2022cs} reports the results of the evaluation in terms of precision, recall, and f-measure. The CS-KG pipeline outperforms all the other tools yielding an overall f-measure of 0.76.         \\ \hline
Curation & No information \\ \hline
Schema &
  The schema is described in \url{https://scholkg.kmi.open.ac.uk/cskg/documentation.php} and \url{https://scholkg.kmi.open.ac.uk/cskg/ontology}.  \\ \hline
Modalities         & Text         \\ \hline
\end{tabularx}
\end{table*}

\begin{table*}[h!]
\begin{tabularx}{\textwidth}{|>{\hsize=.32\hsize}X|>{\hsize=1.68\hsize}X|}
\multicolumn{2}{l}{\textbf{(d) \textit{Quality}}} \\ \hline
\makecell[c]{\textbf{Card}}      & \makecell[c]{\textbf{Description}}     \\ \hline
Accuracy           &  The CS-KG publication \cite{dessi2022cs} in Sections 4.2 and 4.3 describe in detail the automatic methods used to handle the syntactic and semantic errors.                 \\ \hline
Coverage           &      67M statements, 24M entities, 219 relations (\url{https://scholkg.kmi.open.ac.uk/})    \\ \hline
Coherency          &    The CS-KG publication \cite{dessi2022cs} in Section 4.4 describes in detail the automatic methods used to handle the error triples produced. There is no mention for constrain violations or other errors.              \\ \hline
Completeness & No information \\ \hline
Succinctness       &     (i) SPARQL endpoint support (https://scholkg.kmi.open.ac.uk/sparql/), (ii) human-readable labels are provided, (iii) URIs, and (iv) the contact information is available at https://scholkg.kmi.open.ac.uk.          \\ \hline
\end{tabularx}
\end{table*}

\begin{table*}[h!]
\begin{tabularx}{\textwidth}{|>{\hsize=.32\hsize}X|>{\hsize=1.68\hsize}X|}
\multicolumn{2}{l}{\textbf{(e) \textit{Licensing}}} \\ \hline
\makecell[c]{\textbf{Card}}      & \makecell[c]{\textbf{Description}}     \\ \hline
License          & CS-KG is licensed under a Creative Commons Attribution 4.0 International License.    \\ \hline
Publisher           & Team: Danilo Dessì, Francesco Osborne, Diego Reforgiato Recupero, Davide Buscaldi, Enrico Motta. This work is a collaboration of Scholarly Knowledge Modeling, Knowledge Media Institute, The Open University, University of Cagliari, FIZ Karlsruhe, Université Sorbonne Paris Nord.       \\ \hline
Funding & No information \\ \hline
Policy \& Regulation & CS-KG is aligned with the initiative of the Knowledge Graph Construction W3C Community Group for producing benchmarks, resources, and tools to support the semi-automatic generation of knowledge graphs from documents. \\ \hline
\end{tabularx}
\end{table*}

\begin{table*}[t]
\begin{tabularx}{\textwidth}{|>{\hsize=.32\hsize}X|>{\hsize=1.68\hsize}X|}
\multicolumn{2}{l}{\textbf{(f) \textit{Transparency}}} \\ \hline
\makecell[c]{\textbf{Card}}      & \makecell[c]{\textbf{Description}}     \\ \hline
Data Balance           &  Statistics for the first version of CS-KG are provided in Section 5 of the publication~\cite{dessi2022cs}. \\ \hline
Explainability         & No information            \\ \hline
Limitations       & The main limitation of CS-KG is that it was produced with a fully automatic methodology, so the specific statements are not revised by humans, as in manually crafted KGs. The authors also working on developing an entity linking tool for automatically mapping documents (e.g., articles, patents, educational material) to entities and statements in CS-KG. Finally, they plan to further extend the ontology and the entity typing process, in particular by providing a more granular categorization of entity types.
\\ \hline
Fairness Metrics       & No information            \\ \hline
Applications           &  CS-KG can support a variety of intelligent services, such as advanced literature search, document classification, article recommendation, trend forecasting, hypothesis generation, and many others~\cite{dessi2022cs}.              \\ \hline
\end{tabularx}
\end{table*}

\end{document}